\documentclass{article}  
\usepackage{spadre2008}
\usepackage{graphicx}
\usepackage{amssymb}
\newcommand{\dd}  { {\textrm d}}

\frompage{000} \topage{000}                                              %|
\def\rightmark{Prospect of Rapidity Asymmetry and Nuclear Modifications}
\def\leftmark{G.G. Barnaf\"oldi et al.}
\title{Prospect of Rapidity Asymmetry and \\ Nuclear Modifications} 
\authors{
{G.G. Barnaf\"oldi$^{1,2}$, A. Adeluyi$^1$, G. Fai$^{1}$, P. L\'evai$^2$, G. 
Papp$^3$ %
}\\[2.812mm]
{\normalsize
\hspace*{-8pt}$^1$ Center for Nuclear Research, Department of Physics \\ 
Kent State University, Kent OH 44242, USA\\[0.2ex] 
\hspace*{-8pt}$^2$ MTA KFKI RMKI Research Institute for Particle 
and Nuclear Physics\\ 
P.O. Box 49, Budapest 1525, Hungary\\[0.2ex]
\hspace*{-8pt}$^3$ Department for Theoretical Physics, E\"otv\"os University\\ 
P\'azm\'any P\'eter s\'etany 1/A, Budapest 1117, Hungary
}}
\abstract{In asymmetric heavy ion collisions like $dA$ or $pA$, particle
production yields are different in the forward ($d$- or $p$-side) and
backward ($A$-side) rapidity directions. The rapidity distribution reflects 
the geometry and phase-space distribution of nuclear matter. These 
properties may depend on the time evolution of the collision. Due to the 
smallness of the backward-forward differences, the rapidity asymmetry 
factor can be useful to quantify nuclear modification effects, like
e.g. shadowing and the EMC effect. Our work is a survey of the nuclear modification 
factor and the rapidity asymmetries at RHIC energies. We analyze the rapidity 
dependence and the strength of the  nuclear effects. We focus on the high 
transverse momentum region, and make predictions for the role of nuclear 
modifications and rapidity asymmetries for future experimental 
measurements at increasing absolute values of rapidity.}
\keyword{pQCD, intrinsic $k_T$, pion production, Cronin effect}
\PACS{24.85.+p, 13.85.Ni, 13.85.Qk}
\begin{document}
\maketitle
\setcounter{page}{1}
%%%%%%%%%%%%%%%%%%%%%%%%%%%%%%%%%%%%%%%%%%%%%%%%%%%%%%%%%%%%%%%%%%%%%%%%%% 

\section{Introduction}
\label{intro}

Measuring nuclear modifications is not only interesting by itself, but 
is crucial to the understanding of the hard-probe signature of the quark-gluon 
plasma (QGP) formed in high energy heavy ion 
collisions. The modifications can be determined experimentally by different 
methods in a wide kinematical region: (i) the nuclear modification factor, 
$R^h_{AA'}(p_T)$ can magnify the deviation caused by collective 
nuclear effects relative to 'simple' nucleon-nucleon collisions; (ii)
the (pseudo)rapidity asymmetry, $Y^h_{Asym}(p_T)$, measures differences 
of the hadron spectra between backward and forward directions.
The deviation from unity of the above two quantities originates 
in the geometrical and nuclear properties of the colliding system.

Measuring $R^h_{AA'}(p_T)$ and $Y^h_{Asym}(p_T)$ was an important  
task for past and present experimental collaborations~\cite{e706,star}, 
and is going to remain at the center of attention
of future experiments.%~\cite{exps}. 
In this paper we would like to 
emphasize the connection between the two quantities and present recent 
results~\cite{Ade08,Cole07,bgg:qm04} for upgraded detectors at RHIC and future 
experiments at the LHC.

%%%%%%%%%%%%%%%%%%%%%%%%%%%%%%%%%%%%%%%%%%%%%%%%%%%%%%%%%%%%%%%%%%%%%%%%%% 
\section{Quantifying Modifications in the Nuclear Medium}
\label{nuclmod}

The nuclear modification measures the effect of the collective nuclear
forces in a $pA$ or $AA'$ collision, compared to an average nucleon-nucleon
collision. The deviation can be measured as 
a ratio of the given hadron ($h$) spectra per nucleon in  
$AA'$ to the $NN$ collisions. The nuclear modification factor 
%is given on a  linear scale, and it 
can be defined for any given pseudorapidity:
\begin{equation}
R^h_{dA}(p_T, \eta) = \left. \frac{1}{\langle N_{bin}\rangle} \cdot
\frac{E_h \dd^3\sigma_{dA}^{h}/\dd^3 p_T}
{E_h \dd^3\sigma_{pp}^{h}/\dd^3 p_T} \right| _{\eta}   \,\, .
\label{rdau}
\end{equation}
Nuclear effects can make $R^h_{dA}(p_T, \eta)$ grater or smaller than 1,
representing an enhancement or suppression, respectively, relative to the 
$NN$ hadron spectra. 

In asymmetric collisions, hadron production at forward rapidities
may be different from what is obtained at backward
rapidities. It is thus of interest to study ratios of particle
yields between a given pseudorapidity value and its negative  
in these collisions. The pseudorapidity asymmetry $Y_{Asym}(p_T)$ is 
defined for a hadron species $h$ as
\begin{equation}
Y^h_{Asym}(p_T) = \left. E_h\frac{\dd ^3 \sigma_{AB}^{h}}{\dd^3 p_T} \right|_{\eta<0} 
 \left/ 
\left. E_h\frac{\dd ^3 \sigma_{AB}^{h}}{\dd^3 p_T} \right|_{\eta>0} \right. \,\, .
\label{yasym}
\end{equation}

Let us consider the ratio of the backward and forward nuclear modification 
factors in $dAu$ collisions for species $h$:
\begin{eqnarray}
R^h_{\eta}(p_T) &=& \frac{R^h_{dAu}(p_T,\eta<0)}{R^h_{dAu}(p_T,\eta>0)}= \nonumber \\  
&=& \frac{\langle N^{\eta>0}_{bin} \rangle}{ \langle N_{bin} ^{\eta<0} \rangle } 
\cdot \frac{E_h \dd^3\sigma_{dAu}^{h}/\dd^3 p_T |_{\eta<0}}
{E_h \dd^3\sigma_{pp}^{h}/\dd^3 p_T |_{\eta<0}} \left/  
\frac{E_h \dd^3\sigma_{dAu}^{h}/\dd^3 p_T |_{\eta>0}}
{E_h \dd^3\sigma_{pp}^{h}/\dd^3 p_T |_{\eta>0}} \right. \,\, , 
\label{y-r:eq}
\end{eqnarray}
where $\langle N_{bin} \rangle$ is the average number of binary 
collisions in the various impact-parameter bins, and it is given by the
thickness function of the Glauber model. 
Therefore, 
$\langle N^{\eta>0}_{bin} \rangle =\langle N^{\eta<0}_{bin} \rangle$.
Furthermore, the $pp$ rapidity distribution is symmetric around $y=0$.
Thus, if the same backward and forward (pseudo)rapidity ranges are taken
in both directions (i.e. $ |\eta_{min}| \leq |\eta| \leq |\eta_{max}| $),
then the $pp$ yields cancel in eq.~(\ref{y-r:eq}) and one obtains
that the ratio defined in eq. (\ref{y-r:eq}) is identical to
the pseudorapidity asymmetry eq. (\ref{yasym}):
\begin{equation}
Y^h_{Asym}(p_T)= R^h_{\eta}(p_T)=\frac{R^h_{dAu}(p_T,\eta<0)}{R^h_{dAu}(p_T,\eta>0)} \,\, .
\label{mainyrdau}
\end{equation}

%%%%%%%%%%%%%%%%%%%%%%%%%%%%%%%%%%%%%%%%%%%%%%%%%%%%%%%%%%%%%%%%%%%%%%%%%% 

\section{Nuclear Modification at Forward and Backward Rapidities}
\label{sec:r_dau}

We analyzed several types of shadowing parameterizations such as 
HKN~\cite{hkn}, FGS~\cite{fgs}, EKS~\cite{eks98}, HIJING~\cite{hij}, and 
the most recent EPS08~\cite{eps08}.  We choose the two latter 
parameterizations, which are characterized by the largest nuclear effects.  
Both have strong suppression at lower $x$, where we augmented HIJING with 
multiple scattering corresponding to the saturated Cronin model with  
$C=0.35$ (GeV/c)\textsuperscript{2} as in Refs.~\cite{FLP01,Yi02}. 
The EPS08 is based on the earlier EKS, however the authors included not 
only $eA$ data, but the latest results from the RHIC $dAu$ 
experiment~\cite{BRAHMS}. Inclusion of the RHIC data results in  
strong suppression at small $x$ values. In parallel a large 
anti-shadowing appears at around $x\approx 0.2$, due to the 
normalization. 
\vspace*{-.5cm}
\begin{figure}[!htb]
\begin{center}
\includegraphics[width=13.5cm, height=6.5cm, angle=0]{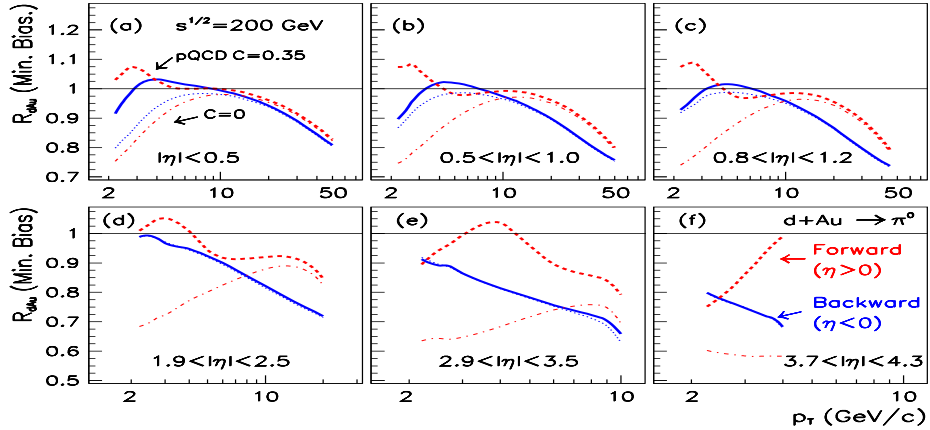}
\end{center}
\vspace*{-0.5cm}
\caption[]{(Color online.) Nuclear modifications to the forward and backward
direction based on HIJING~\cite{hij} shadowing parameterization. (See text 
for details.)}
\label{hij:womulti}
\end{figure}

Applying these shadowing parameterizations we calculated  
$R^h_{dAu}(p_T,\eta)$ in wide ranges of $\eta$ from backward to forward. 
Results are presented on Fig.~\ref{hij:womulti} for HIJING. 
Forward and backward calculations are plotted with 
{\sl (red) dashed} and {\sl (blue) solid} respectively. We also plotted using 
{\sl thin dotted and dash-dotted lines} on Fig.~\ref{hij:womulti} HIJING 
calculations for $R^h_{dAu}(p_T,\eta)$ without multiple scattering indicated 
by '$C=0$'. The differences on Fig.~\ref{hij:womulti} between the 
pQCD calculations with and without multiple scattering diverge between 
$\eta > 0$ and $\eta<0$. In the backward direction the lack of scattering 
centers in the $d$ nucleus limits the possible collisions, thus the two
curves merge as $\eta$ increases. In the forward direction, without 
multiple scattering HIJING gives a suppression at small $p_T$ related to 
low $x$ values. The deviation is growing as $\eta$ increases. Note 
the lack of experimental data in the backward direction. The opening 
with increasing $|\eta|$ between the curves without and with multiple 
scattering suggests an $\eta$-dependent multiple 
scattering similarly to Ref.~\cite{AntoniScurek}.
%
%\vspace*{-.5cm}
\begin{figure}[!htb]
\begin{center}
\includegraphics[width=13.5cm, height=6.5cm, angle=0]{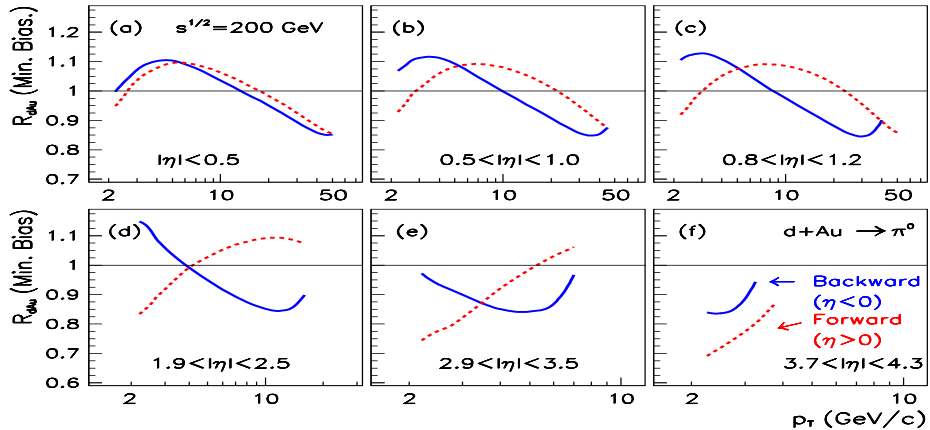}
\end{center}
\vspace*{-.5cm}
\caption[]{(Color online.) Nuclear modifications to the forward and backward
direction based on EPS08~\cite{eps08} shadowing parameterization. (See text 
for details.)}
\label{eps:nomulti}
\end{figure}

On Fig.~\ref{eps:nomulti} we calculated the $R^h_{dAu}(p_T,\eta)$ 
applying the EPS08 shadowing parameterizations.  {\sl Solid (blue)} 
lines represent the backward, {\sl dashed (red)} lines are for the forward
calculations. In the EKS/EPS08 frameworks the effect of multiple scattering 
(Cronin e.g. in Ref.~\cite{Cron79}) is modeled with a strong anti-shadowing peak at 
around $x\approx 0.2$. At small $\eta$ forward and backward have 
similar slopes, but at $\eta \gtrsim 1.5-2.0$ this trend flips over
due to reaching the steep positive slope from the new low-$p_T$ 
RHIC $dAu$ data. High $\eta$ and low $p_T$ data are fitted well with EPS08, 
but at high $p_T$ the anti-shadowing peak overestimates the data around 
midrapidity~\cite{bgg:qm08v2}.

\section{Analyzing Rapidity Asymmetry Data}
\label{rapasym}

\begin{figure}[!htb]
\begin{center}
\includegraphics[width=13.5cm, height=6.5cm, angle=0]{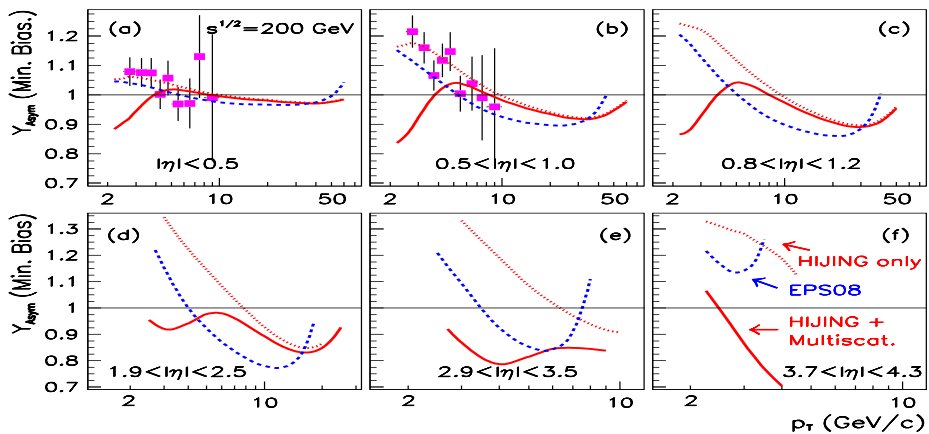}
\end{center}
\vspace*{-.5cm}
\caption[]{(Color online.) Rapidity asymmetry calculated with HIJING~\cite{hij} 
and EPS08~\cite{eps08} shadowing parameterizations. (See ~\cite{star} for data and 
text for details.)}
\label{y_daur}
\end{figure}

Fig.~\ref{y_daur} shows the calculated rapidity asymmetry, $Y^h_{Asym}(p_T)$ 
in the same negative/positive $\eta$ ranges previously presented for 
$R^h_{dAu}(p_T,\eta)$. Based on eq.~(\ref{mainyrdau}) this is the ratio 
of the nuclear modification factors in the proper backward and forward 
rapidity ranges. Since STAR~\cite{star} has published data on pseudorapidity 
asymmetry, this gives a nice opportunity to compare different models 
with measured values reflecting the nuclear modifications in the backward
direction. EPS08, plotted with {\sl (blue) dashed lines}, agrees well with 
the data in a wide $p_T$ range. HIJING itself ({\sl dotted red curves})
shows a similar trend, but inclusion of multiple scattering with $C=0.35$ 
(GeV/c)\textsuperscript{2} turns over the slope below $p_T \lesssim 4-5$ 
GeV/c at higher $\eta$ ({\sl solid red lines}). 
Similarly to the $R^h_{dAu}(p_T,\eta)$ studies in Sec.~\ref{sec:r_dau}, this 
suggests that the efficiency of multiple scattering should decrease going 
beyond midrapidity.            
%
%\vspace*{-.5cm}
%

%%%%%%%%%%%%%%%%%%%%%%%%%%%%%%%%%%%%%%%%%%%%%%%%%%%%%%%%%%%%%%%%%%%%
\section{Conclusions}\label{concl}

We analyzed the nuclear modification factor, $R^h_{dAu}(p_T,\eta)$ and 
the (pseudo)rapidity asymmetry, $Y^h_{Asym}(p_T)$ in a wide rapidity range. 
We showed that these physical properties are related to each other: both 
reflect the geometry of the collisions and nuclear modifications. We 
found the equivalence between them in a general formula. This relation 
led us to test our calculated nuclear modification factors on backward 
hadron production, which has not been measured directly yet.
% at high $\eta$ and $p_T$ values.  
  
Nuclear modifications at midrapidity follow an $x$ scaling in a wide 
kinematical range as we presented in Refs.~\cite{bgg:qm08v2,bgg:qm08v1}. 
The pQCD improved parton model agrees with experimental data from CERN 
SPS up to RHIC energies~\cite{bgg:qm04,Yi02}. Several shadowing parameterizations 
and nuclear PDFs were tested to get the best $\chi^2$ with the 
experimental data. We found that off-midrapidity the $x$-scaling 
is violated using standard shadowing parameterizations. 

To correct this problem, a rapidity dependent multiple scattering 
would be desirable, which is related to the path length (number of 
scattering centers) of a particle in the system. This would 
be equivalent to a strongly $b$-dependent inhomogeneous shadowing
parameterization.

%%%%%%%%%%%%%%%%%%%%%%%%%%%%%%%%%%%%%%%%%%%%%%%%%%%%%%%%%%%%%%%%%%%%
\section*{Acknowledgments}
Discussions with Dieter R\"ohrich [BRAHMS] and Jamie Dunlop [STAR] are
thankfully acknowledged. Our work was supported in part by Hungarian 
OTKA PD73596, T047050, NK62044, and IN71374, by the U.S. Department of 
Energy under grant U.S. DOE DE-FG02-86ER40251. 
%%%%%%%%%%%%%%%%%%%%%%%%%%%%%%%%%%%%%%%%%%%%%%%%%%%%%%%%%%%%%%%%%%%%
%\section*{Note(s)} 
%\begin{notes}
%\item[a]
%E-mail: bgergely@rmki.kfki.hu
%\end{notes}
%%%%%%%%%%%%%%%%%%%%%%%%%%%%%%%%%%%%%%%%%%%%%%%%%%%%%%%%%%%%%%%%%%%%

\vfill\eject
\end{document}